\documentclass[letterpaper,12pt]{JHEP3}
\usepackage{bm}
\usepackage{amsmath}
\usepackage{yfonts}
\newcommand{\qn}{\textswab{q}}
\newcommand{\wn}{\textswab{w}}
\newcommand{\<}{\langle}
\renewcommand{\>}{\rangle}
\renewcommand{\d}{\partial}
\newcommand{\N}{{\cal N}}
\renewcommand{\O}{\hat{\mathcal{O}}}
\renewcommand{\k}{\bm{k}}
\newcommand{\x}{\bm{x}}

\renewcommand{\Re}{{}\,\mathrm{Re}\,}
\renewcommand{\Im}{{}\,\mathrm{Im}\,}
\newcommand{\gYM}{g_{\mathrm{YM}}}
\newcommand{\AdS}{\mathrm{AdS}}
\def\ofo{ { {}_2 \! F_1 }}

\title{Minkowski-space correlators in AdS/CFT correspondence: recipe 
and applications}

\author{Dam T.~Son and Andrei O.~Starinets\\
Institute for Nuclear Theory, University of Washington,
Seattle, WA 98195-1550, USA\\
Emails: \email{son@phys.washington.edu, starina@phys.washington.edu}
}
\preprint{INT-PUB 02-34}

\date{April 2002}

\abstract{
We formulate a prescription for computing 
Minkowski-space correlators from AdS/CFT correspondence.
This prescription is shown to give the correct retarded propagators
at zero temperature in four dimensions, as well as at finite temperature
in the two-dimensional conformal field theory dual to the BTZ 
black hole.
Using the prescription, we
calculate the Chern-Simons
diffusion constant of the finite-temperature $\N=4$ supersymmetric
Yang-Mills theory in the strong coupling limit.
We explain why the quasinormal frequencies 
of the asymptotically 
AdS background correspond to the poles of the retarded 
Green's function of the boundary conformal field theory.
}
\begin{document}
\section{Introduction}
\label{sec:intro}
The main prescription of the anti-de Sitter/conformal field theory
(AdS/CFT) correspondence~\cite{Maldacena}, which allows  
one to compute the correlators in the boundary CFT, was originally 
formulated for Euclidean signature~\cite{GKP,Witten}, 
and has been successfully used ever since.
Working in Euclidean space is a common and convenient practice
which usually does not lead to any restrictions since the results obtained 
in Euclidean AdS/CFT can then be analytically continued to 
Minkowski space, if desired.

In many cases, however, the ability to extract the 
Lorentzian-signature AdS/CFT results directly from gravity 
is indispensable. Many interesting properties of gauge theories at 
finite temperature 
and density, most notably the response of the thermal ensemble to small
perturbations that drive it out of equilibrium,
can only be learned from real-time Green's functions.
Even in the absence of a gravity dual of QCD,
the AdS/CFT correspondence
may provide helpful insights into properties 
of thermal gauge theories at strong coupling.
% in particular, 
%in situations when the traditional perturbative approach 
%faces serious technical difficulties and 
%lattice simulations cannot be used in principle.

In principle, one may try to avoid the Minkowski formulation of the
AdS/CFT correspondence altogether by working only with the Euclidean
version and using analytic properties of the Green's
functions to find the real-time propagators.  In practice, however,
such a method is not useful, because analytic continuation to the
Minkowski space is possible only when the Euclidean correlators are
exactly known for {\it all} Matsubara frequencies.  Since gravity
calculations in non-extremal backgrounds usually involve
approximations of some sort (such as the high- and low-temperature
expansions used in ref.~\cite{Policastro:2001yb}), exact expressions
involving all Matsubara frequencies are normally beyond reach.  It is
therefore crucial to have a prescription allowing one to compute the
Minkowski correlators directly from gravity.

Subtleties of the Lorentzian-signature AdS/CFT correspondence
are well known 
\cite{Balasubramanian:1998de,Balasubramanian:1999ri,Balasubramanian:1998sn}.
In this paper we shall not try to give a formulation of the Minkowski
AdS/CFT in a form as general as the
Euclidean version.  Our goal is to formulate a working recipe, whose
justification and full understanding would hopefully emerge in future research.
The simple prescription that will be given is
sufficient for the computation of the retarded propagator, and hence
all other two-point correlators, but does not allow one to calculate
higher-point correlation functions.  We defer the treatment of
higher-point correlators to future work.

The paper is structured as follows. 
In section~\ref{sec:corr} we review the properties of the 
Euclidean and Minkowski thermal Green's functions.
We formulate our recipe in section~\ref{sec:prescr} and 
illustrate it on a simplest zero-temperature example. 
We compute the Minkowski-space correlators of the operator dual to the 
dilaton in $\AdS_5$, and show that they agree with the analytic continuation 
of the classic Euclidean result. 
%The massive case is treated in 
%appendix~\ref{sec:zero_temperature}.
In section~\ref{sec:BTZ} another nontrivial check of the prescription
is performed:
 we compute the retarded Green's functions in the two-dimensional CFT dual to 
the Ba\~nados-Teitelboim-Zanelli (BTZ) black hole background.
%The imaginary part of the retarded Green's 
%functions is related to the absorption cross-section, and thus can be 
%shown to agree with the known field theory result. We check that for 
%the discrete complex values of frequency (corresponding to Matsubara
%frequency in Euclidean thermal field theory) the retarded Green's functions 
%computed from gravity coincide with the momentum space Euclidean 
%correlators of the field theory. 
We show that the result coincides with the analytic continuation
from Euclidean space.
We also find that the poles of the retarded 
Green's functions correspond precisely to the quasinormal frequencies of 
the general BTZ black hole, and we explain why this is to be expected 
in general in the framework of our recipe.
Section~\ref{sec:CS} is devoted the computation of the Chern-Simons 
diffusion rate in the the strongly coupled ${\cal N}=4$ 
supersymmetric Yang-Mills (SYM) theory.  
The result
cannot be tested independently, and is considered 
as a prediction.
%for the strongly coupled ${\cal N}=4$ SYM theory.
%We compute the Chern-Simons diffusion rate and show that in the strong 
%coupling regime it is proportional to ??? and is independent of the 
%gauge coupling.
Section \ref{sec:concl} contains the conclusions and outlook.
The Appendices are devoted to various calculations outside
the main line of the paper.

The prescription developed in this paper is further applied to the
hydrodynamic regime of the $\N=4$ SYM theory in ref.~\cite{hydro}.

\section{Minkowski thermal correlators}
\label{sec:corr}

Since the main topics of this paper is the computation of thermal
Green's functions from gravity, let us review the definition and general
properties of different thermal Green's functions.
Consider a quantum field theory at finite temperature.  Let $\O$ be an
arbitrary local operator, which we assume, for definiteness, to be
bosonic (the fermionic case is completely analogous). 
In Minkowski space,
the retarded propagator for $\O$ is defined by
\begin{equation}
  G^R(k) = -i \int\!d^4x\, e^{-ik\cdot x}\, \theta(t)\,
  \< [\O(x), \O(0)] \>
\label{ret}
\end{equation}
(we use the --+++ metric convention, and $t\equiv x^0$).  
The advanced propagator is defined similarly,
\begin{equation}
  G^A(k) = i \int\!d^4x\, e^{-ik\cdot x}\, \theta(-t)\,
  \< [\O(x), \O(0)] \>\,.
\end{equation}
From the definitions can see that $G^R(k)^*=G^R(-k)=G^A(k)$.  If the
system is P-invariant, then ${\rm Re}\,G^{R,A}$ are even functions of
$\omega\equiv k^0$ and ${\rm Im}\,G^{R,A}$ are odd functions of $\omega$.
Instead of
$G^R$ and $G^A$, one can consider other correlation functions of $\O$
in thermal equilibrium. One example is the symmetrized Wightman function,
\begin{equation}
  G(k) = {1\over2} \int\!d^4x\, e^{-ik\cdot x}
  \< \O(x)\O(0) + \O(0)\O(x)\>\,.
\end{equation}
All other correlation functions can be expressed via $G^R$, $G^A$ and $G$.
For example, the Feynman propagator is
\begin{equation}
  G^F(k) = -i\int\!d^4x\, e^{-ik\cdot x}\, \<T\O(x)\O(0)\> =
  {1\over2}[G^R(k)+G^A(k)] - i G(k)\,.
\end{equation}

From the spectral representations of $G^R$ and $G$, one can relate
$G(k)$ with the imaginary part of the retarded propagator,
\begin{equation}
  G(k) = -\coth{\omega\over 2T}\, {\rm Im}\, G^R(k)\,.
  \label{fluctdiss}
\end{equation}
%Equation~(\ref{fluctdiss}) is the fluctuation-dissipation theorem.
Therefore, if $G^R(k)$ is known, all other correlators are easy to
compute.  In particular,
\begin{equation}\label{GFGRT}
  G^F(k) = {\rm Re}\, G^R(k) + i \coth\frac\omega{2T}\, 
           {\rm Im}\, G^R(k)\,.
\end{equation}
At zero temperature ($T\to0$), eq.~(\ref{GFGRT}) reduces to a simple formula,
\begin{equation}\label{GFGRT=0}
  G^F(k) = {\rm Re}\, G^R(k) + i\,{\rm sgn}\, \omega \, {\rm Im}\,G^R(k)
  \,, \qquad T=0 \,.
\end{equation}
Taking the limit $\omega\to0$ in eq.~(\ref{fluctdiss}), one
obtains another useful formula,
\begin{equation}
  G(0,\k) = -\lim_{\omega\to0}{2T\over\omega}{\rm Im}\, G^R(k) =
  2iT {\d\over\d\omega}G^R(\omega,\k)\Bigl|_{\omega=0}\,.
  \label{GdGR}
\end{equation}

The multitude of Minkowski-space Green's functions is in contrast with
the situation in Euclidean space, where
one normally deals only with the Matsubara propagator,
\begin{equation}
  G^E(k_E) = \int\!d^4 x_E\, e^{-ik_E\cdot x_E} \< T_E\O(x_E) \O(0) \>\,.
\end{equation}
Here $T_E$ denotes Euclidean time ordering.  The Matsubara propagator is
defined only at discrete values of the frequency $\omega_E$.  For bosonic $\O$
these Matsubara frequencies are multiples of $2\pi T$.

The Euclidean and Minkowski propagators are closely related.
The retarded propagator $G^R(k)$, as a function of $\omega$, can
be analytically continued to the whole upper half plane and, moreover,
at complex values of $\omega$ equal to $2\pi i Tn$, reduces to the
Euclidean propagator,
\begin{equation}
  G^R(2\pi iTn, \k) = - G^E (2\pi Tn, \k)\,.
\label{compR}
\end{equation}
Analogously, the advanced propagator, analytically continued to the
lower half plane, is equal to the Matsubara propagator at the points
$\omega=-2\pi i Tn$,
\begin{equation}
  G^A(-2\pi iTn, \k) = -G^E (-2\pi Tn, \k)\,.
\end{equation}
In particular, by putting $n=0$, one finds that
$G^R(0,\k)=G^A(0,\k)= - G^E(0,\k)$.

\section{Minkowski AdS/CFT prescription}
\label{sec:prescr}

\subsection{Difficulties with Minkowski AdS/CFT}
Let us first recall the formulation of the AdS/CFT correspondence in
Euclidean space.  For definiteness, we shall talk about the
correspondence between the strongly coupled $\N=4$ SYM theory and
classical (super)gravity on $\AdS_5\times\mathrm{S}^5$.  
The Euclidean version of
the metric of the latter has the form
\begin{equation}
  ds^2 = {R^2\over z^2} (d\tau^2 + d\x^2 + dz^2) + R^2 d\Omega_5^2\,,
  \label{extr-met}
\end{equation}
which is a solution to the Einstein equations.  In the AdS/CFT
correspondence, the four-dimensional quantum field theory lives on the
 boundary of the $\AdS_5$ space at $z=0$.  Suppose that a
bulk field $\phi$ is coupled to an operator $\O$ on the
boundary in such a way that the interaction Lagrangian is
$\phi\, \O$.  In this case, the AdS/CFT correspondence is formally
stated as the equality
\begin{equation}
  \< e^{\int_{\d M}\phi_0\O} \> = e^{-S_{\rm cl}[\phi]}\,,
  \label{EuclAdSCFT}
\end{equation}
where the left-hand side is the generating functional for correlators of 
 $\O$ in the
boundary field theory (i.e., ${\cal N}=4$ $SU(N)$ 
SYM theory at large $N$ and large
't Hooft coupling $g^2 N$), and the exponent on the right-hand side is
the action of the classical solution to the equation of motion for
$\phi$ in the bulk metric with the boundary condition
$\phi|_{z=0}=\phi_0$.

The metric~(\ref{extr-met}) corresponds to the zero-temperature field
theory.  To compute the Matsubara correlator at finite temperature, one
has to replace it by a non-extremal one,
\begin{equation}
  ds^2 = {R^2\over z^2} \Bigl( f(z) d\tau^2 + d\x^2 +
  {dz^2\over f(z)}\Bigr) + R^2 d\Omega_5^2\,,
  \label{nonextr-met}
\end{equation}
where $f(z) = 1-z^4/z_H^4$ and $z_H=(\pi T)^{-1}$, and $T$ is the Hawking
temperature.  The Euclidean time coordinate $\tau$ is periodic,
$\tau\sim\tau+T^{-1}$, and $z$ runs between 0 and $z_H$.

One can try to formally write the Minkowski version of the AdS/CFT
 as the equivalence
\begin{equation}\label{MinkAdSCFT}
  \< e^{i \int_{\partial M} \phi_0 {\cal O}}\> = e^{iS_{\rm cl}[\phi]}\,.
\end{equation}
However, an immediate problem arises.  In the Euclidean version, 
the classical solution $\phi$  is uniquely
determined by its value $\phi_0$ at the boundary $z=0$ and
the requirement of regularity at the horizon $z=z_H$.
Correspondingly, the Euclidean correlator obtained by using the
correspondence is unique.  In contrast, in the Minkowski space, the
requirement of regularity at the horizon is insufficient; to select a
solution one needs a more refined boundary condition there.  This
problem is well known~\cite{Balasubramanian:1998de},
and is thought to reflect
the multitude of real-time Green's functions (Feynman, retarded,
advanced) in finite-temperature field theory.  Previous 
discussion of the Lorentzian signature AdS/CFT appears also in 
refs.~\cite{Balasubramanian:1999ri,Balasubramanian:1998sn,%
Danielsson:1998wt,Ryang:1999xm}.

One boundary condition at the horizon stands out from the physical
point of view.  It is the incoming-wave boundary condition, 
where waves can only
travel to the region inside the horizon of the black branes, but cannot be
emitted from there.  We may suspect that this boundary
condition corresponds to the retarded Green's function, while the
outgoing-wave boundary condition gives rise to the advanced Green's
function.  However, even after fixing the boundary condition at
the horizon, it is still problematic to get eq.~(\ref{MinkAdSCFT}) 
to work.

To see where the problem lies,
first we notice that the
AdS part of the metric~(\ref{nonextr-met}) is of the form
\begin{equation}
  ds^2 = g_{zz} dz^2 + g_{\mu\nu} (z) d x^\mu  d x^\nu\,.
\label{background_0}
\end{equation}
Let us consider fluctuations of a scalar (e.g., the dilaton or the
axion) on this background.  
The action of the scalar reads
\begin{equation}
  S =  K \int\!d^4x\,\int\limits_{z_B}^{z_H}\!dz\, \sqrt{-g} 
  \left[ g^{zz} 
(\partial_z \phi )^2 + g^{\mu\nu}\d_\mu\phi\d_\nu\phi 
  +m^2\phi^2\right]\,,
\label{action_0}
\end{equation}
where $K$ is a normalization constant (for the dilaton
$K=-\pi^3R^5/4\kappa_{10}^2$, $\kappa_{10}$ is the ten-dimensional 
gravitational
constant) and $m$ is the scalar mass.  
The limit of the $z$-integration
is between the boundary $z_B$ and the horizon $z_H$ (for the
metric~(\ref{nonextr-met}), $z_B=0$).

The linearized field equation for $\phi$,
\begin{equation}\label{phi-eq}
  \frac1{\sqrt{-g}} \d_z (\sqrt{-g}\,g^{zz}\d_z\phi) 
  + g^{\mu\nu}\d_\mu\d_\nu\phi - m^2\phi =0 \,,
\end{equation}
has to be solved with fixed boundary condition at $z_B$.  The solution is
\begin{equation}
  \phi (z,x) = \int\!{d^4 k\over (2\pi)^4}\,
 e^{ik\cdot x} f_k(z)\phi_0(k)\,,
\end{equation}
where $\phi_0(k)$ is determined by the boundary condition,
\begin{equation}
  \phi(z_B,x) = \int\!\frac{d^4k}{(2\pi)^4}\, e^{ik\cdot x} \phi_0(k)\,,
\end{equation}
and $f_k(z)$ is the solution to the mode equation,
\begin{equation}\label{mode-eq}
  \frac1{\sqrt{-g}} \d_z (\sqrt{-g}\,g^{zz}\d_z f_k)
  - (g^{\mu\nu}k_\mu k_\nu+m^2)f_k = 0 \,,
\end{equation}
with unit boundary value at the boundary, $f_k(z_B)=1$ and satisfying the 
incoming-wave boundary condition at $z=z_H$.

The action on shell (i.e., for classical solutions) reduces to the 
surface terms
\begin{equation}\label{Sboundary}
S = \int\!{d^4k\over (2\pi )^4}\, 
  \phi_0(-k) {\cal F} (k,z) \phi_0(k) \Big|^{z=z_H}_{z=z_B}\,,
\end{equation}
where
\begin{equation}
 {\cal F} (k,z) =  K \sqrt{-g} g^{zz}f_{-k}(z)\d_z f_k(z)\,.
\label{flux_0}
\end{equation}

Now, if one try to make the identification~(\ref{MinkAdSCFT}), the
Green function can be obtained by taking second derivative of the
classical action with respect to the boundary value of $\phi$.  From
eq.~(\ref{Sboundary}), this Green's function is
\begin{equation}\label{Gboundary}
  G(k) = -{\cal F}(k,z)\Bigl|_{z_B}^{z_H} 
         - {\cal F}(-k ,z)\Bigr|_{z_B}^{z_H}\,.
\end{equation}
One can see, however, that this quantity is completely real, and hence 
cannot be a candidate to the retarded Green's function, which is in general
complex.  Indeed, by noticing that
$f^*_k(z) = f_{-k}(z)$ (this is because $f^*_k(z)$ is also a solution to
eq.~(\ref{mode-eq}) and also satisfies the incoming-wave boundary condition),
one sees that 
the  imaginary part of ${\cal F} (k,z)$ is proportional a
conserved flux,
\begin{equation}
 \Im {\cal F} (k,z) = {K \over 2 i} \sqrt{-g}\,g^{zz}
 \left[ f_k^* \d_z f_k  - f_k \d_z f_k^* \right]\,,
\label{flux_im}
\end{equation}
and thus  $\Im {\cal F} (k,z)$ is independent of the radial 
coordinate, $\d_z \Im  {\cal F} =0$.  Therefore, in each term in the
right-hand side of eq.~(\ref{Gboundary}), the imaginary part at the
horizon $z=z_H$ completely cancels out the imaginary part at the boundary 
$z=z_B$.

We should clarify one issue with the boundary term at the horizon.  It
is commonly thought that this term is oscillating and hence averaged
out to zero.  However, this is not the case if the incoming-wave
boundary condition is consistently maintained.  We shall see that
explicitly below in section~\ref{sec:zeroT}.

We can try to avoid this problem by throwing away the contribution
from the horizon, keeping only the boundary term at $z=z_B$.  However,
now the imaginary parts of the two terms in eq.~(\ref{Gboundary})
cancel each other: from the reality of the field equation one can show
that ${\cal F}(-k,z)={\cal F}^*(k,z)$.  Therefore, the resulting
$G(k)$ is still real.  It seems that by differentiating the classical
action one cannot get the retarded Green's function, which is,
in general, complex.

\subsection{The Minkowski prescription}

We circumvent the difficulties mentioned above
by putting forward the following conjecture
\begin{equation}\label{G2F}
  G^R(k) = -2{\cal F}(k,z)\Bigl|_{z_B}\,.
\end{equation}
which seems rather natural but, for reasons explained above, does not
follow strictly from an identity of the type~(\ref{MinkAdSCFT}).
The justification for eq.~(\ref{G2F}) is that it works in all
cases where independent verification is possible.

%where the Green's function $G$ is related to the correlator 
%via $i G =  <{\cal O} {\cal O}>$ and $ {\cal F}_{\partial M}$ 
%is the {\it boundary} term,
%\begin{equation}
% \phi_0 {\cal F}_{\partial M}\phi_0 
% = - K \lim_{z\rightarrow z_B}%\sqrt{-g} g^{zz}\phi_{-k}(z)\d_z\phi_k(z)\,.
%\end{equation}
% Thus,
%the proportionality between $G$ and  $ {\cal F}_{\partial M}$ 
%is fixed as $G = -2 {\cal F}_{\partial M} $.

Our prescription for computing the retarded (advanced) Green's functions
is thus formulated as follows: 

1. Find a solution
to the mode equation~(\ref{mode-eq}) with the following properties:

$\bullet$ The solution equals to 1 at the boundary $z = z_B$;

$\bullet$  For timelike momenta, the solution has an asymptotic expression 
corresponding to the incoming (outgoing) wave
at the horizon. For spacelike momenta, the solution is regular at the horizon.

2. The retarded Green's function is then given by 
$G = -2 {\cal F}_{\partial M}$, where ${\cal F}$ is given in 
eq.~(\ref{flux_0}).
Only the contribution from the boundary has to be taken.
%We stress that in (\ref{prescription}) the only contribution 
%to be taken into account is the one coming from the boundary $\partial M$
%where the dual theory is defined. 
Surface terms coming from the horizon or, 
more generally, from the ``infrared'' part of the background geometry 
(corresponding to the ``position of the branes'') must be dropped. 
This part of the metric influences the correlators only through the 
boundary condition imposed on the bulk field $\phi$.

Since the imaginary part of ${\cal F} (k,z)$ is independent of the
radial coordinate, $\Im G^{R,A}$ can be computed by evaluating $\Im
{\cal F} (k,z)$ at any convenient value of $z$.  In particular, it can
be computed at the horizon.

In order to verify that the prescription indeed gives the 
retarded propagator $G^R$, one should
compute $G^R$ in theories where it is known
from other methods.  Below we perform a check for zero-temperature
field theory; further checks are done in section~(\ref{sec:BTZ}) and
ref.~\cite{hydro}.

\subsection{Example: Zero-temperature field theory}
\label{sec:zeroT}
As an example, we use the prescription (\ref{G2F}) 
to compute the retarded (advanced) two-point function of the 
composite operators ${\cal O} = \frac14 F^2$ at zero temperature.
In this case the action (\ref{action_0}) is that for 
a minimally coupled massless scalar on the background (\ref{extr-met}).
%(\ref{background_0}) with $g_{zz}=1/z^2$, $g_{\mu\nu} = \eta_{\mu\nu}/z^2$,
%$\eta_{\mu\nu} = \mbox{diag} (-1,1,1,1)$. 
The horizon is at $z_H=\infty$ and
the boundary at $z_B = \epsilon \rightarrow 0$.
The mode equation reads
\begin{equation}
  f_k ''(z) - {3\over z} f_k '(z) - k^2 f_k (z) = 0\,.
\label{eom_0}
\end{equation}
For spacelike momenta, $k^2>0$, the calculation is identical to the one
for the Euclidean case~\cite{GKP} (see also 
appendix~\ref{sec:zero_temperature}), 
the only difference being the extra minus sign in front of the 
Lorentzian signature action. We obtain
\begin{equation}
 G_{R} (k) =  {N^2 k^4\over 64 \pi^2} \ln{k^2}\,,\;\;\;\;\; k^2 >0\,.
\label{spacelike_0}
\end{equation}
For timelike momenta, we introduce
$q=\sqrt{-k^2}$.  The solution to eq.~(\ref{eom_0}) satisfying
boundary conditions outlined above is given by
\begin{equation} 
   f_k (z)   \;=\;
%   \cases{
\begin{cases} 
 \displaystyle{  { z^2 H_{2}^{(1)}(qz)
\over \epsilon^2  H_{\nu}^{(1)}(q\epsilon ) }\,,}
                          &  \omega > 0,  \cr
           \noalign{\vskip 4pt}
          \displaystyle{  { z^2 H_{2}^{(2)}(qz)
\over \epsilon^2  H_{2}^{(2)}(q\epsilon ) }\,, }
 &  \omega < 0.   \cr
\end{cases}
%         }
%   \label{f_grav}
\end{equation}
Notice that $f_{-k} = f^*_k$.
Computing  (\ref{flux_0}) and using eq.~(\ref{G2F}), we get
\begin{equation}
 G^R (k) =  {N^2 k^4\over 64 \pi^2} (\ln{k^2} - 
 i\pi\, \mbox{sgn} \, \omega)\,.
\label{timelike_0}
\end{equation}
Combining eqs.~(\ref{spacelike_0}) and  (\ref{timelike_0}),
we obtain the retarded Green's function,
\begin{equation}
 G^R (k) =  {N^2 k^4\over 64 \pi^2}( \ln{|k^2|} - 
i\pi\,  \theta (-k^2)\, \mbox{sgn} \, \omega)\,.
\label{ret_massless}
\end{equation}
Using the asymptotics of the Hankel functions for $z\rightarrow \infty$,
one can check that ${\cal F}(k,z)$ is not zero at large $z\to\infty$.
Moreover, it is purely imaginary in this limit, and is
equal to $i N^2 k^4 \mbox{sgn} \, \omega /128 \pi$ $=
\Im {\cal F}(k,\epsilon)$, as it should be in view of 
the flux conservation (\ref{flux_im}). Thus, the 
imaginary part of the Green's function can be computed 
independently using the asymptotics 
of the solution at the horizon.

By using the relation between the Feynman and retarded propagators at
zero temperature, eq.~(\ref{GFGRT=0}), we find
\begin{equation}
  G^F (k) =  {N^2 k^4\over 64 \pi^2} (\ln{|k^2|} - 
  i\pi\,  \theta (-k^2))\,,
\end{equation}
which can be obtained from the Euclidean correlator,
\begin{equation}
 G_{E} (k_E) = - {N^2 k_E^4\over 64 \pi^2} \ln{k_E^2}\,,
\end{equation}
by a Wick rotation.  Thus, our prescription reproduces the correct answer
for the retarded Green's function at zero temperature.

The Euclidean and Minkowski correlators corresponding to massive 
scalar modes in $\AdS_5$ are computed in 
appendix~\ref{sec:zero_temperature}.

\section{Retarded correlators in 2d CFT and the BTZ black hole}
\label{sec:BTZ}

In this section,
we use the prescription~(\ref{G2F}) in the 
general BTZ black hole background
to  compute the retarded Green's functions of the dual two-dimensional CFT.
The result is then analytically continued to complex frequencies, and is
shown to reproduce the Matsubara correlators known from field theory.
The calculation is technically quite involved, but is 
nevertheless instructive since
it is the first nontrivial check of our prescription at
finite temperature.
As a by-product, we demonstrate that the gravitational 
quasinormal frequencies in the asymptotically 
AdS background correspond to the poles of the retarded 
correlator in the dual CFT on the boundary,
and thus provide a quantitative description of the 
system's return to thermal equilibrium.

\subsection{Gravity calculations}

The non-extremal BTZ black hole \cite{Banados:wn,Banados:1992gq} 
is a solution of 
the $2+1$-dimensional Einstein equations with a cosmological
 constant $\Lambda = - 1/l^2$. The metric is given by
\begin{equation}
ds^2 = - { (\rho^2 -\rho_+^2) (\rho^2 -\rho_-^2)\over l^2 \rho^2} dt^2 +
\rho^2 \left( d\varphi - {\rho_+\rho_-\over l \rho^2} dt\right)^2 +
 { l^2 \rho^2\over  (\rho^2 -\rho_+^2) (\rho^2 -\rho_-^2)} d\rho^2\,,
\label{btz_metric}
\end{equation}
where $\varphi$ is periodic with the period $2\pi$ and $\rho_{\pm}$
 correspond to the positions of the inner/outer horizons.
Introducing new variables $\mu$, $x_{\pm}$, 
with 
\begin{equation}
\begin{split}
\rho^2 &= \rho_+^2 \cosh^2\mu -  \rho_-^2 \sinh^2\mu \,,\\
 x^{\pm} &= \pm {\rho_{\pm} t\over l} \mp \rho_{\mp} \varphi\,,
\end{split}
\end{equation}
we can write the metric as
\begin{equation}
ds^2 = l^2 d\mu^2 - \sinh^2\mu (dx^+)^2 + \cosh^2\mu (dx^-)^2\,.
\end{equation}
In terms of $z = \tanh^2 \mu$, the action of a massive scalar in the 
BTZ background reads
%\begin{eqnarray}
\begin{equation}\label{action}
\begin{split}
S &= - \eta 
 \int\! dz\, dx^+\, dx^-\, {l\over 2 (1-z)^2}\biggl[ {4z(1-z)^2\over l^2}
\left( \partial_z \phi \right)^2 - {1-z\over z} 
\left( \partial_+ \phi \right)^2 \\
&\quad+ (1-z) \left( \partial_- \phi \right)^2 + m^2 \phi^2 \biggr]\,,
\end{split}
\end{equation}
%\end{eqnarray}
where $\eta$ is the normalization constant.

The mode functions $f_{k_+,k_-}(z)$ are defined so that
\begin{equation}
 \phi (x^{\pm}, z) = 
 \int\! {d k_+\, dk_-\over (2\pi )^2} \,
  e^{-i ( k_+ x^+ + k_- x^-)} f_{k_+,k_-}(z) \phi_0(k_+,k_-)\,,
\end{equation}
and satisfy the equation
\begin{equation}
\partial_{zz}^2 f_{k_+,k_-} + {1\over z} \partial_z f_{k_+,k_-} +
\left[ {l^2 k_+^2\over 4 z^2 (1-z)} - {l^2 k_-^2\over 4 z(1-z)}
 - {m^2 l^2\over 4 z (1-z)^2}\right]f_{k_+,k_-} = 0 \,.
\label{btz_equation}
\end{equation}
On shell, the action~(\ref{action}) reduces to the boundary terms,
\begin{equation}
S = \int\!  {d k_+ dk_-\over (2\pi )^2} \,
   \phi_0(-k_+,-k_-) \left[ {\cal F} (k_+, k_-, z_B) -
 {\cal F} (k_+, k_-, z_H)\right] \phi_0(k_+,k_-)\,,
\end{equation}
where $z_B = 1$ corresponds to the boundary at infinity ($\rho = \infty$),
$z_H = 0$ is the location of the outer horizon ($\rho = \rho_+$). 
The function $\mathcal{F}$ is given by 
\begin{equation}
 {\cal F} (k_+, k_-, z) = - \eta {2 z\over l} f_{k_+,k_-}^*\partial_z
 f_{k_+,k_-}\,. 
\end{equation}
According to the Minkowski AdS/CFT prescription, 
the retarded/advanced Green's functions are determined by
\begin{equation}
G^{R,A} (k_+, k_-) = -2 [ \lim_{z\rightarrow z_B}{\cal F} (k_+, k_-, z)]\,, 
\label{btz_prescription}
\end{equation}
where the solution $f_{k_+,k_-}(z)$ is normalized to $1$ at $z = 1 -\epsilon$
and represents a purely incoming (outgoing) wave at the horizon $z=0$. 
Square brackets indicate that the contact terms are ignored.
The imaginary part of $G^{R,A}$, due to the flux conservation, can be computed
either at the boundary or at the horizon,
\begin{equation}
\Im G^{R,A} (k_+, k_-) = 
-2 [ \lim_{z\rightarrow z_B}\Im {\cal F} (k_+, k_-, z)] =
-2 [ \lim_{z\rightarrow z_H}\Im {\cal F} (k_+, k_-, z)]\,. 
\label{im_part_0}
\end{equation}

The solution to eq.~(\ref{btz_equation}) can be written as
\begin{equation}
f_{k_+,k_-}(z) = 
z^{\alpha} (1-z)^{\beta} \epsilon^{-\beta} (1-\epsilon )^{-\alpha}
 {\ofo \left( a, b; c; z\right)\over 
\ofo \left( a, b; c; 1-\epsilon \right)}\,,
\label{solution}
\end{equation}
where
\begin{equation}
 a, b = {l(k_+ \mp k_-)\over 2 i} + \beta\,,
 \qquad c = 1 + 2\alpha\,,
\end{equation}
and the indices $\alpha$, $\beta$ are given by
\begin{equation}
  \alpha_{\pm} = \pm i l k_+/2\,,\qquad
  \beta_{\pm} = {1\over 2 }\left( 1 \pm \sqrt{1 + l^2 m^2}\right) = 
{\Delta_{\pm}\over 2}\,,
\end{equation}
$\alpha_-$ corresponds 
to the incoming wave at the horizon, $\alpha_+$ to the outgoing one,
and
$\Delta_{\pm}$ is the conformal weight of the boundary CFT operator.
The range of $\Delta_+$ is $[1,\infty)$, while the range of 
 $\Delta_-$ is $(0, 1]$, where zero corresponds to
 the unitarity bound in $d=3$. Here we shall
 consider only the  $\Delta_+$ branch.
Accordingly,\footnote{In AdS/CFT, 
if a supergravity scalar behaves near the boundary 
as $f \sim A \epsilon^{d-\Delta_+} + B \epsilon^{\Delta_+}$, 
$A$ is interpreted as the source of a dimension-$\Delta_+$ operator.}
we choose $\beta = \beta_-$ in eq.~(\ref{solution}).

%\subsection{Retarded/Advanced correlation functions}
%\label{sec:corr}

\subsubsection{The full retarded Green's function}
\label{sec:full_green}
Near the boundary, i.e., at $z = 1-\epsilon$, the derivative of the solution
(\ref{solution}) takes the form
\begin{equation}
 f_{k_+,k_-}'(z) |_{z=1-\epsilon} = -{\beta \over \epsilon} + 
 {a b \ofo \left( a+1, b+1; c+1; 1-\epsilon\right)\over 
c \ofo \left( a, b; c; 1-\epsilon \right)}\,,
\label{near_eps}
\end{equation}
The retarded Green's function is then computed using the prescription
(\ref{btz_prescription}),
\begin{equation}
G^R (k_+,k_-) = {4\eta \over l }\left[ {a b\over c}
 \lim_{\epsilon\rightarrow 0}  
{\ofo \left( a+1, b+1; c+1; 1-\epsilon\right)\over 
 \ofo \left( a, b; c; 1-\epsilon \right)}\right] \,.
\label{green_eps}
\end{equation}
Some care should be taken when computing the limit in (\ref{green_eps})
for  $1 - 2\beta_+ = c - a - b = - n$, with $n=0,1,2,...$, 
i.e., for  $\Delta_+ = 2\beta_+ \neq n+1$. 
In this case the hypergeometric function is degenerate
and contains logarithmic terms. Accordingly, integer and non-integer
 $\Delta_+$ need to be considered separately.

\subsubsection{Non-integer $\Delta_+$}
For $\Delta_+ = 2\beta_+ \neq n+1$, where $n=0,1,2,...$,
direct calculation gives
%\begin{eqnarray}
\begin{equation}\label{full_green}
\begin{split}
 G_{R}(p_+ , p_-) & =  -\,
  {2\, \eta \, \epsilon^{2\beta_+ -2}
\over  \pi\,  l\,
  \Gamma^2 ( 2 \beta_+  -1)\sin{2\pi\beta_+} }
\left| \Gamma \left( \beta_+ +  {i p_+\over 2 \pi T_L }\right) 
 \Gamma \left( \beta_+ +  {i p_-\over 2 \pi T_R }\right)\right|^2
  \\ 
  &\quad \Biggl[ \cosh{\left(  {p_+\over 2 T_L } -
 {p_-\over 2 T_R}  \right)  } -\cos{2\pi\beta_+} 
\cosh{\left(  {p_+\over 2 T_L } + {p_-\over 2 T_R} 
 \right)  }  
  \\ 
  &\quad+ i \sin{2\pi\beta_+} 
\sinh{\left(  {p_+\over 2 T_L } + {p_-\over 2 T_R}  \right)  }\Biggr]\,,
\end{split}
\end{equation}
%\end{eqnarray}
where  $T_{L,R} = (\rho_+ \mp \rho_-)/2\pi$, and 
the momenta $p_{\pm}$ are related to $k_{\pm}$ and $\omega,k$ 
as follows
\begin{equation}
p_{+} = \pi T_{L}  \left( k_+ + k_-\right) l = {\omega - k\over 2}\,, \qquad \,
p_{-} = \pi T_{R}  \left( k_+ - k_-\right) l = {\omega + k\over 2}\,.
\end{equation}
%
%\begin{equation}
%p_{-} = \pi T_{R}  \left( k_+ - k_-\right) l = {\omega + k\over 2}\,.
%\end{equation}
%
%\begin{equation}
%p_{\pm} = \pi T_{L,R} l \left( k_+ \pm k_-\right) = {\omega \mp k\over 2}\,.
%\end{equation}
%
\subsubsection{Integer $\Delta_+$}
In the case of integer conformal dimension, one has to further distinguish
between  $\Delta_+$ being an odd or even integer.
In the latter case, i.e., for  
$h_L = h_R = \beta_+ = \Delta_+/2 = n + 1$, $n=0,1,2,...$,
we get
%\begin{eqnarray}
\begin{equation}\label{a33}
\begin{split}
 G_{R}(p_+ , p_-) & = 
 {4 \eta \epsilon^{2 n}\over l \pi^2 \Gamma^2 (2 n +1)} \left|
\Gamma \left( 1 + n + {i p_+\over 2 \pi T_L} \right)
\Gamma \left( 1 + n + {i p_-\over 2\pi T_R} \right) \right|^2 
 \\
&\quad\sinh { p_+\over 2  T_L}  \sinh { p_-\over 2  T_R} 
\left[ \psi \left( 1 + n  - {i p_+\over 2 \pi T_L}\right) +
 \psi \left( 1 + n  - {i p_-\over 2 \pi T_R}\right)\right]\,.
\end{split}
\end{equation}
%\end{eqnarray}
In particular, the following simple result is obtained 
for $h_L = h_R = 1$ ($\Delta_+ = 2$): 
\begin{equation}
 G_{R}(p_+ , p_-) = {\eta  p_+ p_-\over \pi^2 l T_L T_R} 
\left[ \psi \left( 1 - {i p_+\over 2\pi T_L}\right) +
 \psi \left( 1 - {i p_-\over 2\pi  T_R}\right)\right]\,.
\label{green_1}
\end{equation}
The imaginary part of  (\ref{green_1}),
\begin{equation}
 \Im G_{R}(p_+ , p_-) = - {\eta p_+ p_-\over 2 \pi l  T_L T_R}
 \left( \coth{{p_+\over 2 T_L}} +  \coth{{ p_-\over 2 T_R}}\right)\,,
\end{equation}
coincides with the  appropriate limit of eq.~(\ref{im_green}) below.
For odd integer values of $\Delta_+$, i.e., for 
$h_L = h_R = \beta_+ = \Delta_+/2 = n + 1/2$, $n = 1,2,...$,  we get 
%\begin{eqnarray}
\begin{equation}\label{a34}
\begin{split}
 G_{R}(p_+ , p_-) & = 
 {4 \eta \epsilon^{2 n -1}\over l \pi^2 \Gamma^2 (2 n)} \left|
\Gamma \left( {1\over 2} + n + {i p_+\over 2 \pi T_L} \right)
\Gamma \left( {1\over 2}
 + n + {i p_-\over 2\pi T_R} \right) \right|^2 \\
&  \cosh { p_+\over 2  T_L}  \cosh { p_-\over 2  T_R} 
\left[ \psi \left( {1\over 2} + n  - {i p_+\over 2 \pi T_L}\right) +
 \psi \left( {1 \over 2} + n  - {i p_-\over 2 \pi T_R}\right)\right]\,.
\end{split}
\end{equation}
%\end{eqnarray}
Finally, for 
 $\beta_+ = 1/2$ ($\Delta_+ = 1$) we have
\begin{equation}
 G_{R}(p_+ , p_-) = {4\, \eta \over l \epsilon \ln^2 \epsilon}
\left[ \psi \left(  {1\over 2} - {i p_+\over 2\pi T_L}\right) +
 \psi \left( {1\over 2} - {i p_-\over 2\pi  T_R}\right)\right]\,.
\label{green_1/2}
\end{equation}
The imaginary part of the obtained Green's functions coincides with 
eq.~(\ref{im_green}) below.
More crucially, one can check that the only 
singularities of (\ref{full_green}), (\ref{a33}), (\ref{a34}) and
 (\ref{green_1/2}) are simple poles located 
in the lower half-plane of complex $\omega$ at
\begin{subequations}\label{freq_12}
\begin{eqnarray}
\omega_n^{(L)} &=& k - i 4\pi T_L ( h_L + n)\,, \;\;\; n=0,1,2,\dots\,,
 \\
\omega_n^{(R)} &=& - k - i 4\pi T_R ( h_R + n)\,,  \;\;\; n=0,1,2,\dots\,.
\end{eqnarray}
\end{subequations}
No singularity is located in the upper half plane, as
expected for the retarded Green's function.

\subsubsection{A note about the imaginary part}
Notice that if we are interested only in the  
imaginary part of the retarded propagator, the calculations are much
simpler because it can be done
by using 
the solution~(\ref{solution})
with $\alpha = \alpha_-$, $\beta = \beta_-$ and taking the 
limit $z\rightarrow z_H$ in eq.~(\ref{im_part_0}).  The result reads
\begin{equation}
\Im G_{R} = -\, {\cal C}(\epsilon, h_{L,R})
 \sinh{\left( {p_+\over 2 T_L }+ {p_-\over 2 T_R}\right)}
\left| \Gamma \left( h_L +  {i p_+\over 2 \pi T_L }\right) 
 \Gamma \left( h_R +  {i p_-\over 2 \pi T_R }\right)\right|^2\,,
\label{im_green}
\end{equation}
where 
$h_L = h_R = \beta_+$. The 
normalization constant is 
\begin{equation} 
  {\cal C}(\epsilon, h_{L,R})   \;=\;
%   \cases{
\begin{cases} 
\displaystyle{  {2\, \eta \, \epsilon^{h_L + h_R -2}
\over  \pi\,  l\,
  \Gamma ( 2 h_L  -1) \Gamma ( 2 h_R  -1)}\,,}
                          &   \Delta_+ > 1,  \cr
           \noalign{\vskip 4pt}
          \displaystyle{{2\,  \eta\,  \over \pi\, l\, 
          \epsilon \ln^2\epsilon}\,, 
}
 &   \Delta_+ = 1  .   \cr
\end{cases}
%         }
   \label{normalization}
\end{equation}
This result has been known before from the absorption 
calculations~\cite{Gubser:1997cm}.

%The absorption cross-section is related to the imaginary part of the retarded
%Green's function at zero spatial momentum via
%\begin{equation}
%\sigma_{abs} \sim {1\over \omega} \Im G^R (\omega, 0)\,,
%\end{equation}
%where we have ignored the $\omega$-independent coefficient.
%From  (\ref{im_green}) we obtain
%\begin{equation}
%\sigma_{abs} \sim { \sinh{{\omega\over 2 T_H}}\over \omega }
% \left| \Gamma \left( h_L +  {i \omega\over 4 \pi T_L }\right) 
% \Gamma \left( h_R +  {i \omega\over 4 \pi T_R }\right)\right|^2\,,
%\label{abs}
%\end{equation}
%where $T_H$ is the Hawking temperature ($2/T_H = 1/T_R +1/T_L$).
%eq.~(\ref{abs}) agrees precisely with the field theory result 
%\cite{Gubser:1997cm}. {\bf Why compare absorption with field theory?}

\subsection{Comparison with the two-dimensional finite-temperature CFT}
Now let us show that the retarded Green's function found from gravity
coincides with the field-theory result.
One starts from the coordinate-space expression for 
the Euclidean finite-temperature correlation function of the 
local operator ${\cal O}$ with conformal dimensions 
$(h,\bar{h})$~\cite{Cardy:rp}
%is the correlator  
%in the two-dimensional CFT on the cylinder with circumference 
%$\beta = 1/T$, and is given by 
\begin{equation}
G^E(\tau,x) = \langle {\cal O}(w,\bar{w})
 {\cal O}(0,0)\rangle  
=  {C_{\cal{O}}\, (\pi T)^{4h+4\bar{h}}   \over 
\sinh^{2h}[\pi T w] \, \sinh^{2\bar{h}}[\pi T\bar{w}] 
}\,,
\label{cardy}
\end{equation}
where $C_{\cal{O}}$ is a normalization constant and
$w=x+ i\tau$, with $x$ running from $-\infty$ to $+\infty$, and $\tau$ 
being periodic with the period $1/T$.

To compare with the results obtained from gravity, we need to
compute the Fourier transform of (\ref{cardy})
for $h=\bar{h}$,
obtain the momentum-space Matsubara correlator, and
% defined at discrete
%values of frequency $\omega_E = 2\pi n T$. We 
compare the result with the 
retarded Green's functions obtained from gravity in 
section~\ref{sec:full_green}.
%taken at complex values of $\omega$ equal to 
%$2\pi i T n$ (with integer $n$) and show that formula (\ref{compR}) holds.
We make the comparison only for $T_L = T_R$ and for the integer values of 
$\Delta = 2h = 2\bar{h}$.
%\footnote{The first simplification is made for a 
% technical reason, the second one 
%reflects our inability to compute the Fourier 
%transform in the case of non-integer $\Delta$.}
We need to compute the following integral,
\begin{equation}
G^E(\omega_E =2\pi n T, k) = \int\limits_{0}^{1/T}\!d\tau\!\int\! dx\, 
e^{i\omega_E \tau}\, e^{i k x}\, G^E(\tau,x)\,.
\label{fourier_E}
\end{equation}
It is convenient to write $G^E(\tau,x)$ in the form
\begin{equation}
 G^E(\tau,x)
= {C_{\cal{O}} (2\pi T)^{2\Delta}\over (\cosh{2\pi T r} - 
\cos{2\pi T\tau})^{\Delta}}\,.
\end{equation}
% Note that (\ref{fourier_E}) is
%invariant under $n\rightarrow -n$ due to periodicity of  
%$G^E(\tau,x)$ in $\tau$.
We shall first take the integral over $\tau$ and then over $x$.
Introducing $z= \exp{(-i 2\pi T\tau)}$ and $r=|x|$, the integral becomes
\begin{equation}
G^E(\omega_E, k) = {(-1)^{2\Delta}2 (2\pi T)^{2\Delta}C_{\cal{O}}\over T}
 \int\limits_{0}^{\infty}\!dr \cos{k r}  
\oint\limits_{|z|=1} {dz\over 2\pi i} {z^{\Delta - n -1}\over (z-z_+)^{\Delta}
 (z-z_-)^{\Delta}}\,,
\end{equation}
where $z_{\pm} = \exp{(\pm 2\pi T r)}$.
The integral over $z$ is computed by evaluating the residues.  The result
reads
\begin{equation}
G^E(\omega_E, k) = {2 (2\pi T)^{2\Delta}C_{\cal{O}}\over T}
{\Gamma (\Delta - n)\over \Gamma (\Delta)}
\sum\limits_{m=0}^{\Delta - 1} {\Gamma (2\Delta - m -1)\over m!
\Gamma (\Delta -m )\Gamma (\Delta -m-n ) } J_{n,m}(k)\,,
\label{btz_res}
\end{equation}
where $J_{n,m}(k)$ are the following integrals that can be computed
explicitly~\cite{prudnikov},
\begin{equation}
J_{n,m}(k) =  \int\limits_{0}^{\infty}\! dr\, { e^{2\pi T r (\Delta + n)}
 \cos{k r} \over (e^{4\pi T r }-1)^{2\Delta - m -1}} = 
{(-1)^{m-2\Delta}\Gamma (s)\over 8 \pi T} \left[ 
{\Gamma ( z-s)\over \Gamma (z)} + {\Gamma ( \bar{z}-s)\over \Gamma (\bar{z})}
\right]\,,
\label{int_j}
\end{equation}
and $s = 2-2\Delta +m$, and
\begin{equation}
z = 2 - {3\Delta\over 2} + m + {n\over 2} - {i k\over 4\pi T}\,.
\end{equation}
In general $J_{n,m}(k)$ diverges, and the result on the right-hand side of 
eq.~(\ref{int_j}) should be understood as a limit of the regularized 
expression; after taking the limit, the (divergent) 
contact terms should be ignored.

As an example, consider the simplest case $\Delta = 1$. Taking the limit
$s\rightarrow 0$ in eq.~(\ref{int_j}), we get
\begin{equation}
J_{n,0}(k) = - {1\over 8\pi T} \Biggl[ \psi (z) + \psi (\bar{z})\Biggr]\,.
\end{equation}
Therefore, we obtain
\begin{equation}
G^E(2\pi n T, k) = -\pi \, C_{\cal{O}} 
\left[ \psi \left( {1+n\over 2} -
{i k \over 4\pi T} \right) + \psi \left( {1+n\over 2} +
{i k \over 4\pi T} \right)\right]\,.
\label{green_E_1/2}
\end{equation}
Identifying the prefactors as $4\eta/l \epsilon \ln^2\epsilon = 
\pi C_{\cal{O}}$ one observes that the retarded Green's function
(\ref{green_1/2}) taken at $\omega = 2\pi i T n$ equals 
(with a minus sign) to the expression given by 
eq.~(\ref{green_E_1/2}), as expected
on general grounds in eq.~(\ref{compR}),
and therefore the field theory and gravity results are consistent.

For $\Delta =2$, eqs.~(\ref{btz_res}) and (\ref{int_j}) give
\begin{equation}
G^E(2\pi n T, k) = \pi \, C_{\cal{O}} {(2\pi T n)^2+k^2\over 4}
\left[ \psi \left( 1+ {n\over 2} -
{i k \over 4\pi T} \right) + \psi \left( 1+ {n\over 2} +
{i k \over 4\pi T} \right)\right]\,.
\label{green_E_1}
\end{equation}
Upon identification 
$\pi^3 C_{\cal{O}} = \eta/l T_R T_L$, 
(\ref{green_E_1}) 
coincides with  $-G^R (\omega = 2\pi T n,k)$, where $G^R$ was 
computed in eq.~(\ref{green_1}). One can repeat this procedure 
for any integer $\Delta$.  So we have checked that our prescription
does give the correct retarded Green's functions in two-dimensional
CFT.

\subsection{Quasinormal modes and singularities of $G^R$}

The quasinormal modes for black holes in asymptotically AdS spacetimes
are defined as the solutions to the wave equation 
obeying the ``incoming wave'' boundary condition at the horizon and the 
vanishing Dirichlet condition at the boundary.\footnote{For reviews and
references on quasinormal modes in {\it asymptotically flat}
spacetimes see~\cite{frolov,Kokkotas:1999bd}.} It was suggested in
ref.~\cite{Horowitz:1999jd} that the associated quasinormal frequencies 
are related to the process of thermalization 
in the dual strongly coupled CFT.\footnote{Similar ideas appear in 
ref.~\cite{KalyanaRama:1999zj}. Other works on quasinormal modes in 
asymptotically AdS spacetimes include 
refs.~\cite{Chan:1996yk,Wang:2000gs,% 
Cardoso:2001bb,Wang:2001tk,Moss:2001ga}.}
The inverse relaxation time is given by the imaginary part of the lowest 
quasinormal frequency, which has been computed numerically 
in ref.~\cite{Horowitz:1999jd} for 
AdS black holes in various dimensions.

Due to the simplicity of the BTZ background, 
the quasinormal modes in this case
can be determined analytically~\cite{Birmingham:2001hc,%
Govindarajan:2000vq,Cardoso:2001hn}.
It was first noticed in ref.~\cite{Birmingham:2001pj} that the 
quasinormal frequencies found in ref.~\cite{Birmingham:2001hc}
 correspond to the poles of the 
retarded Green's function describing the relaxation of the boundary CFT 
to thermal equilibrium, as in the
linear response theory.
%\footnote{The condition used 
%in ref.~\cite{Birmingham:2001pj} to
%relate gravity and CFT was the requirement for the flux to vanish at the
%boundary. However, for the quasinormal frequencies the flux
%does not actually vanish at the boundary, it is only regular there.
%In any case, it is not clear from the 
%point of view of AdS/CFT why this condition should be imposed.}

Our calculation of the retarded Green's function confirms this conjecture
explicitly: the poles of $G^R$ (eqs.~(\ref{freq_12}))
are precisely the quasinormal frequencies of the BTZ 
black hole~\cite{Birmingham:2001hc}.

In fact, given the prescription~(\ref{G2F}), 
it is not difficult to see that this should be true in general.
Indeed, suppose $\phi_k(z) = A(k)\phi_1(z) + B(k)\phi_2(z)$ 
is the (super)gravity field whose boundary 
value $\phi_0$ couples to an operator ${\cal O}$ 
in the dual theory in Minkowski space, and suppose further
that $\phi$ obeys the incoming wave boundary condition at the horizon.
The normalized solution used to compute the retarded correlator has the form
$\psi_k(z) = \phi_0(k)  \phi_k(z)/\phi_k(\epsilon)$. Near the boundary
$z=\epsilon\rightarrow 0$ one has $\psi_k(z) = A(k)\, z^{\Delta_-} +\cdots + 
B(k)\, z^{\Delta_+}+\cdots$. Then
\begin{equation}
G^R (k) \sim \psi^* \psi' = (\Delta_+-\Delta_-) {B(k)\over A(k)} 
\epsilon^{\Delta_+-\Delta_--1} + \mbox{contact terms} + O(\epsilon)\,.
\end{equation}
Zeros of $A(k)$ correspond to singularities of the retarded correlator.
But $A(k)=0$ is precisely the vanishing Dirichlet boundary condition
at $z=0$ which defines the quasinormal modes in asymptotically 
AdS spacetimes (recall that the boundary condition at the horizon is the
same for the quasinormal modes and for the solution from 
which $G^R$ is determined). Thus, quite generically, 
quasinormal frequencies are equal to the 
the resonance frequencies 
of the retarded Green's function of the operator ${\cal O}$.
In the linear response theory,
they characterize the time scale of the approach 
to thermal equilibrium in the boundary field theory perturbed by ${\cal O}$.

\section{Thermal $\N=4$ SYM theory and the Chern-Simons diffusion rate}
\label{sec:CS}

We now consider the four-dimensional $\N=4$ SYM theory at
finite temperature.  The thermodynamics of this theory has been
studied in detail by using its gravity dual~\cite{entropy,entropy1}.  
Here, we compute a simple thermal
correlation function by using the Minkowski AdS/CFT prescription.

The quantity we shall be interested in is the Chern-Simons diffusion
rate,
\begin{equation}
  \Gamma = \left(\frac{\gYM^2}{8\pi^2}\right)^2
   \int\!d^4x\, \< \O(x) \O(0) \>\,,
  \label{Gamma}
\end{equation}
where
\begin{equation}\label{OFFdual}
  \O = {1\over4} F^a_{\mu\nu}\tilde F^{a\mu\nu}\,.
\end{equation}
%where $\<\cdots\>$ denote thermal average.  More intuitively, $\Gamma$
%defines the rate at which the Chern-Simons number 
%\begin{equation}
%  \NCS = \cdots
%\end{equation}
%diffuses.  In a volume $V$ and during a time $T$, the mean square of
%the difference between the Chern-Simons numbers at initial and final
%times, $\<[\NCS(t)-\NCS(0)]^2\>$, is equal to $\Gamma VT$.  Hence,
%$\Gamma$ has the dimension of [momentum]$^4$.
This quantity is interesting because it has been intensively studied
at weak coupling.  In the standard model, $\Gamma$ determines the rate
of anomalous baryon number violation at high temperatures.  Its
knowledge is important for electroweak
baryogenesis~\cite{RubakovShaposhnikov}.  At weak coupling, the
parametric behavior of $\Gamma$ is~\cite{ASY,HuetSon,Bodeker,Moore}
\begin{equation}
  \Gamma = \mathrm{const}\cdot (\gYM^2N)^5 \ln \frac1{\gYM^2N}\, T^4\,.
  \label{Bodeker}
\end{equation}
%where $\lambda=\gYM^2N$ is the 't Hooft coupling.  
The overall numerical constant is of nonperturbative nature 
but can be found by a
lattice simulation of a classical field 
theory~\cite{HuMuller,MooreHuMuller,Moorelog}.
No method has been developed to compute $\Gamma$ at strong coupling.

In the original AdS/CFT correspondence, $\O$ is coupled to the bulk
axion, therefore the correlators of $\O$ can be found from gravity.
Indeed, if one finds the retarded propagator $G^R$ for $\O$, then,
according to eq.~(\ref{GdGR}),
\begin{equation}\label{GammaGR}
   \Gamma = - \left(\frac{\gYM^2}{8\pi^2}\right)^2
   \lim_{\omega\to0}{2T\over\omega}\Im 
    G^R(\omega, {\bf 0})\,.
\end{equation}
Therefore, the task of computing $\Gamma$ is reduced to the calculation
of the retarded Green's function of $\O$ at small momenta.

%\subsection{Calculation of the thermal retarded propagator}
%\label{sec:CS_calc}

The coupling of the axion (the Ramond-Ramond 
scalar) $C_0$ to $\O$ is 
determined from the three-brane Born-Infeld 
action~\cite{Klebanov:1997kc,Gubser:1997yh}.  The action for $C_0$ is
the same as for a minimally coupled scalar, eq.~(\ref{action_0}).
It is convenient to use the coordinate $u$ defined as
\begin{equation}
  u = \frac{z^2}{z_H^2}\,.
\end{equation}
The mode equation has the form
\begin{equation}
  f_k '' - {1+u^2\over u (1-u^2)}f_k' + 
  {\wn^2\over  u (1-u^2)^2}f_k - 
  {\qn^2\over  u(1-u^2)}f_k =0\,,
\label{heun_eq}
\end{equation}
where the following notations were introduced,
\begin{equation}
  \wn = \frac\omega{2\pi T}\,,\qquad \qn = \frac {|\k|}{2\pi T}\,.
\end{equation}
Equation~(\ref{heun_eq}) is the Heun differential equation~\cite{heun}
(second-order differential equation with four regular singular points, 
see \cite{bateman,heun_book}).
Global solutions to the Heun equation are unknown.  However, for computing
the Chern-Simons diffusion rate one needs to consider only the regime of
low frequency and momentum, $\omega,|\k|\ll T$.  In this regime, the solution
representing the incoming wave at the horizon is given by
\begin{equation}
  \phi_k(u) = (1-u)^{-i\wn /2} F_k(u)\,,
\end{equation}
where $F_k(u)$  is regular at $u=1$ and can be written as a series,
\begin{eqnarray}
F_k(u) &=& 1 - \biggl({i\wn \over 2}+\qn^2\biggr) \ln { {1+u\over 2}} 
  + O(\wn^2,\,\wn\qn^2,\,\qn^4)\,. 
%{\wn^2\over 8} \left[ \left( \ln{ {1+u\over 2 }} + 8 (1-s^2)\right)
% \ln{ {1+u\over 2 }}  - 4 \mbox{Li}_2 \left( {1-u\over 2}
% \right) \right]\nonumber \\ & +& 
% O(\wn^3)
\end{eqnarray}
Computing the boundary term (\ref{flux_0}) near $u=0$ we obtain
\begin{equation}
  {\cal F}(k,u) = - {\pi^2 N^2 T^4\over 8} \left[ i\wn
  -\qn^2\Bigl(1-\frac1u\Bigr)\right] + O(\wn^2,\,\wn\qn^2,\,\qn^4)\,.
%  + \wn^2 (s^2-1) -  {\wn^2 (s^2-1)\over  u} + 
%  \wn^2 \ln{2}\right) + O(\wn^3, u)
\end{equation}
When $\k=0$, our prescription implies
\begin{equation}
 G^{R/A} (\omega, \bm{0} ) = \mp {i\pi N^2 \omega T^3\over 8 } \left( 1+
 O\left({\omega \over T}\right)  \right)\, \,.
\label{high_T}
\end{equation}
From eq.~(\ref{GammaGR}) one then finds
\begin{equation}\label{Gamma-strongcoup}
  \Gamma= \frac{(\gYM^2N)^2}{256\pi^3} T^4\,,
\end{equation}
which is our final result for the Chern-Simons diffusion rate
at large 't Hooft coupling $\gYM^2N\gg1$.  Note that
(\ref{Gamma-strongcoup}) has the same large $N$ behavior as the
weak-coupling expression
(\ref{Bodeker})
in the 't Hooft limit (that is, of order $N^0$).  
The natural conjecture to make is that the Chern-Simon diffusion rate
is proportional to $T^4$ with a proportionality coefficient dependent
on the 't Hooft coupling,
\begin{equation}
  \Gamma = f_\Gamma(\gYM^2N) T^4
\end{equation}
where $f_\Gamma(x)\sim x^5\ln\frac1x$ for $x\ll1$ and 
$f_\Gamma(x)=x^2/(256\pi^3)$ for $x\gg1$.

Notice that if we considered, instead of $\O=\frac14F\tilde F$ the
operators $\frac14 F^2$, the answer would be the same.  The same
result is also valid for spatial components of the stress-energy
tensor if $\k=0$, since the corresponding gravitational perturbation
obeys the same equation~\cite{Gubser:1997yh}.  For the latter case, we
reproduce, from Kubo's formula, the value for the shear viscosity
$\eta = \frac\pi8 N^2 T^3$ that has been found in
ref.~\cite{Policastro:2001yc}.

The imaginary part of the retarded propagator $G(\omega,\k)$ can also be
computed for very high frequencies and momenta ($\omega,|\k|\gg T$).  This
is done in appendix~\ref{sec:high-freq}.

\section{Conclusions}
\label{sec:concl}
We have formulated a prescription that allows one to compute the
Minkowski retarded Green's function from gravity.  We have demonstrated
that the prescription works for zero-temperature gauge
theory, and for finite-temperature two-dimensional CFT.  By using the
prescription, we found the Chern-Simons diffusion rate for the $\N=4$
SYM theory at finite-temperature.  The prescription allows one to compute
many other quantities which cannot be computed from Euclidean space.
Two examples, the shear viscosity and the R-charge diffusion rate in
thermal $\N=4$ SYM theory, are presented in ref.~\cite{hydro}.

The prescription given here does not allow one to find higher-point
correlation function, and hence it cannot be the final word on
Minkowski gauge theory/gravity duality.  Hopefully, it can be
incorporated in some future more general scheme.

\begin{acknowledgments}
The authors thank I.R.~Klebanov, G.~Policastro, 
and M.~Porrati for
discussions.  This work is supported, in part, by DOE grant No. DOE-ER-41132.
  The work of D.T.S.\ is supported, in part, by the Alfred P.\
Sloan Foundation.
\end{acknowledgments}

\appendix

\section{Zero-temperature correlation functions in \\
Euclidean and Minkowski AdS/CFT}
\label{sec:zero_temperature}

\subsection{Euclidean signature}
To compute the Euclidean two-point function of a CFT operator ${\cal O}$
one uses the AdS/CFT equivalence
\begin{equation}
\langle e^{\int_{\partial M} \phi_0 {\cal O}}\rangle = e^{- S_E[\phi]}\,,
\end{equation}
where $S_E[\phi]$ is the classical (super)gravity action on $M$ and
 $\phi_0$ is the boundary value of the bulk field $\phi$.
At zero temperature $M = \AdS_5\times \mathrm{S}^5$. 
These calculations were first done in ref.~\cite{GKP} and then
appeared in the literature many times.
 We repeat them here
to have an explicit comparison with the Lorentzian version.
The (Euclidean) $\AdS_5$ part of the metric is given by
\begin{equation}
ds^2_{d+1} = {R^2\over z^2} (dz^2 + d\x^2)\,,
\end{equation}
where $\x$ are the coordinates on $R^4$.
The action of  the massive dilaton fluctuation $\phi$ is
\begin{equation}
S_E = {\pi^3 R^8\over 4\kappa^2_{10}} \int\! dz\, d^4 x\, z^{-3}
\left[ (\partial_z \phi )^2 + (\partial_i \phi )^2 + 
{m^2R^2\over z^2}\phi^2\right]\,.
\end{equation}
Using the Fourier representation,
\begin{equation}
  \phi (z, x) = \int\! {d^4 k\over (2\pi )^4}\, e^{i k x}
  f_k (z)\phi_0(k)\,,
\end{equation}
we obtain
\begin{equation}
S_E = {\pi^3 R^8\over 4\kappa^2_{10}} \int\limits_{\epsilon}^{\infty}\!
 dz\! \int\! {d^4 k d^4 k'\over (2\pi )^4 }\, 
 \frac{\delta (k+k')}{z^3}\left[
  \d_z f_k \d_z f_{k'} - k k'f_k f_{k'}
+  {m^2R^2\over z^2}f_k f_{k'}\right]\phi_0(k)\phi_0(k')\,.
\end{equation}
The equation of motion for $f_k (z)$ reads
\begin{equation}\label{fkeq}
  f_k '' - {3\over z}f_k ' - 
  \left( k^2 + {m^2 R^2\over z^2}\right)f_k = 0\,.
\end{equation}
It has a general solution 
\begin{equation}
\phi_k (z) = A z^{2} I_{\nu}(k z) + B z^{2} I_{-\nu} (k z)\,,
\end{equation}
where $\nu = \sqrt{4 + m^2 R^2}$ (or equivalently one can consider 
$I_{\nu}$ and $K_{\nu}$ as two independent solutions). 
We also use the standard notation $\Delta = \nu + 2$ 
for the conformal weight of the operator ${\cal O}$.
The solution regular at $z=\infty$ and equal 1 at
$z=\epsilon$ is given by
\begin{equation}\label{fK}
  f_k (z) = {z^2 K_{\nu}(kz)\over
\epsilon^2 K_{\nu}(k\epsilon )}\,.
\end{equation}
On shell, the action reduces to the boundary terms,
%\begin{eqnarray}
\begin{equation}
\begin{split}
S_E &=  {\pi^3   R^{8}\over 4\kappa^2_{10}}
 \int\! {d^4 k d^4 k'\over (2\pi )^8 }\, 
  (2\pi)^4 \delta^4(k+k')\phi_0(k)\phi_0(k')
  {f_{k'}(z)\d_z f_k (z)\over z^3} \Bigg|_{\epsilon}^{\infty} \\
 &\equiv \int\! {d^4 k d^4 k'\over (2\pi )^8 }\,
\phi_0 (k)\phi_0 (k') {\cal F} (z,k,k') \Big|_{\epsilon}^{\infty} \,.
\end{split}
\end{equation}
%\end{eqnarray}
We note that ${\cal F} (z,k,k')$ is a real function, and 
 ${\cal F} (\infty,k,k') = 0$. 
The two-point function is given by
%\begin{eqnarray}
\begin{equation}
\begin{split}
  \< \O(k)\O(k') \> &=
 Z^{-1}{\delta^2 Z[\phi_0]\over  \delta \phi_0(k)
 \delta \phi_0(k')}\Bigg|_{\phi_0 = 0}  = -2  {\cal F} (z,k,k') 
\Big|_{\epsilon}^{\infty} \\
 & = -  (2\pi)^4\delta^4(k+k') {\pi^3 R^{8}\over  2\kappa^2_{10}} 
  {f_{k'}(z)\d_z
  f_k (z)\over z^3} \Bigg|_{\epsilon}^{\infty}\,.
\end{split}
\end{equation}
%\end{eqnarray}
Direct calculation from eq.~(\ref{fK}) gives
\begin{equation}
  \<\O(k)\O(k')\> = -  {\pi^3 R^{8}\over 2 \kappa^2_{10}}
 \epsilon^{2(\Delta - d)}(2\pi)^4
 \delta^4(k+k') k^{2\nu} 2^{1-2\nu} {\Gamma (1 - \nu )\over \Gamma (\nu)} 
+\cdots \,
\end{equation}
where dots denote terms analytic in $k$ and/or those vanishing 
in the $\epsilon \rightarrow  0$ limit.
Substituting
$\kappa_{10}  = 2\pi^{5/2}R^4/N$~\cite{entropy}
we get
\begin{equation}
  \<\O(k)\O(k')\> = - {N^2\over 8\pi^2} \epsilon^{2(\Delta - 4 )}
  (2\pi)^4\delta^4(k+k') 
  {k^{2\Delta - 4} \Gamma (3-\Delta )\over 2^{2\Delta - 5} 
  \Gamma (\Delta - 2 )}\,.
\end{equation}
 For integer $\Delta$, the propagator is given by
\begin{equation}
\<\O(k)\O(k')\> = - {(-1)^{\Delta}\over (\Delta - 3)!^2}
 {N^2\over 8\pi^2} \epsilon^{2(\Delta - 4 )}
 (2\pi)^4\delta^4 (k+k')  
 {k^{2\Delta - 4}\over 2^{2\Delta - 5}}\ln{k^2}\,, \;\;\;\; 
 \Delta = 2, 3, 4, ...
\end{equation}
where one sets $(-1)!=1$.
Specializing further to the massless case ($\Delta = 4$), we have
\begin{equation}
  \< \O(k)\O(k')\> = - {N^2\over 64 \pi^2}(2\pi)^4\delta^4(k+k')
k^4 \ln{k^2}\,.
\end{equation}
%In the position space this corresponds to 
%\begin{equation}
%< {\cal O}(x)  {\cal O}(y) >  = {3 N^2\over \pi^4 |x-y|^8}\,.
%\end{equation}
%For ${\cal O} = F^2/4g^2$ it gives
%\begin{equation}
%< F^2(x)  F^2(y) >  = {48 (g^2_{YM} N) ^2 \over \pi^4 |x-y|^8}\,.
%\end{equation}

\subsection{Lorentzian signature}
%
%
%The Minkowski version of the AdS/CFT correspondence is 
%formally stated as the equivalence
%\begin{equation}
%\langle e^{i \int_{\partial M} \phi_0 {\cal O}}\rangle = e^{i S[\phi]}\,.
%\end{equation}
%
In the $\AdS_5$ background the action describing the dilaton 
fluctuation reads
%\begin{equation}
%S = - {\pi^3 R^5\over 4 \kappa^2_{10}} \int d^4x
% \int\limits_{\epsilon}^{\infty} dz 
%\sqrt{-g}g^{\mu\nu}\partial_{\mu}\phi \partial_{\nu}\phi  \,,
%\end{equation}
\begin{equation}
S = - {\pi^3 R^8\over 4 \kappa^2_{10}}\! \int d^4x\! 
\int\limits_{\epsilon}^{\infty}\! dz\, 
{dz\over z^3} \left[ (\partial_z \phi )^2 + \eta^{\mu\nu}\partial_{\mu}\phi
\partial_{\nu}\phi + {m^2R^2\over z^2} \phi^2 \right]\,,
\end{equation}
where $\eta_{\mu\nu} = \mbox{diag} (-1,1,1,1)$ and  
$R$ is the $AdS_5$ radius.
%the $AdS_5$ metric is given by 
%\begin{equation}
%ds^2 = { R^2 ( - dt^2 + dx_1^2 + dx_2^2 +dx_3^2 + dz^2)\over z^2}\,,
%\end{equation}
%with $R$ being the $AdS_5$ radius.
Writing the classical solution as 
\begin{equation}
  \phi (z,t,\x) = \int\! {d^4 k\over (2\pi)^4}\,
   e^{-i\omega t + i \k\cdot\x}f_k(z)\phi_0(k)\,,
\end{equation}
the action reduces to the boundary term
\begin{equation}
S = \int\!{d^4k\over (2\pi )^4}\, \Bigl[ {\cal F} (k, \infty ) -
{\cal F} (k, \epsilon )\Bigr]\phi_0(-k)\phi_0(k) \,,
\end{equation}
where
\begin{equation}
 {\cal F} (k,z) = -{\pi^3 R^8\over 4\kappa^2_{10}}{ 
  f_{-k}  \d_z  f_k(z)\over z^3}\,.
\label{flux_pure}
\end{equation}
% Reality condition on $\phi(x)$ implies
%$\phi^*_k = \phi_{-k}$.
The mode function $f_k (z)$ obeys the same equation~(\ref{fkeq})
as in the Euclidean case.  However,
now the form of the solution depends on whether $k$ is spacelike or
timelike.

\subsubsection{Spacelike momenta, $k^2 > 0$}
For spacelike momenta, the solution is real and 
the calculation is identical to that of the 
Euclidean case, except for the additional minus sign coming from the 
difference in definitions of Euclidean and Minkowski actions.
The result is thus given by 
\begin{equation}
G^R (\omega, \k)
 = {N^2 (k^2)^{\Delta - 2} \Gamma (3-\Delta )\epsilon^{2(\Delta -4)}
\over 8\pi^2 2^{2\Delta -5} \Gamma (\Delta -2 )}\,.
\label{green_spacelike}
\end{equation}

\subsubsection{Timelike momenta, $k^2 < 0$}
Introducing  $q=\sqrt{-k^2}=|k|$, the solution to (\ref{fkeq}) satisfying
the ``incoming wave'' boundary condition at $z=\infty$ and normalized to
$1$ at $z=\epsilon$ is
\begin{equation} 
   \phi_k (z)   \;=\;
%   \cases{ 
\begin{cases}
\displaystyle{  { z^2 H_{\nu}^{(1)}(qz)
\over \epsilon^2  H_{\nu}^{(1)}(q\epsilon ) }\,,}
                         \qquad   \omega > 0\,,  \cr
           \noalign{\vskip 4pt}
          \displaystyle{ { z^2 H_{\nu}^{(2)}(qz)
\over \epsilon^2  H_{\nu}^{(2)}(q\epsilon ) }\,, }
  \qquad \omega < 0\,.   \cr
\end{cases}
%         }
%   \label{f_grav}
\end{equation}
Here $\nu = \sqrt{4+m^2R^2}$. 
Using the expansion
\begin{equation} 
z^2 H^{(1,2)}_{\nu}(z) = \pm {1\over \sin{\nu \pi}}
\left[ - {2^{\nu} z^{2-\nu}\over \Gamma (1-\nu )}
\left( 1 + O(z^2)\right) + e^{\mp\nu\pi i}
 { z^{\nu +2}\over 2^{\nu} \Gamma (\nu +1 ) }
  \left( 1 + O(z^2)\right) \right]\,,
\end{equation}
and the prescription~(\ref{G2F}),
 one obtains the retarded Green's function
 for timelike momenta
\begin{equation}
G^R (k)
 = {N^2 (q^2)^{\Delta - 2} \Gamma (3-\Delta )\epsilon^{2(\Delta -4)}
\over 8\pi^2 2^{2\Delta -5} \Gamma (\Delta -2 )}
\left[ \cos{\pi\Delta} - i \sin{(\pi\Delta)} \, \mbox{sgn} \, \omega \right]
\,.
\label{green_timelike}
\end{equation}
Here $\Delta = 2+\nu \neq 2,3,...$.
%Note that this formally corresponds to analytically continuing the 
%Euclidean expression using $(-1)^{\nu} = \exp{-i\nu\pi}$ for $\omega > 0$ and 
% $(-1)^{\nu} = \exp{i\nu\pi}$ for $\omega < 0$.
For integer values of $\Delta$, by taking 
the appropriate limit in eq.~(\ref{green_timelike}), one has
\begin{equation}
G^R (k) =  { N^2  \epsilon^{2(\Delta - 4 )}
\over  8\pi^2(\Delta - 3)!^2  2^{2\Delta - 5} }
  (q^2)^{\Delta - 2}
\left[ \ln{q^2} - i \pi\,  \mbox{sgn}\, \omega\right]\,, \;\;\;\; 
\Delta = 2, 3, 4, ...
\end{equation}
where one sets $(-1)!=1$.

Combining the results for the time- and spacelike momenta, one can write
\begin{subequations}
\begin{equation} 
 \Re G^R (k)     \;=\;
%   \cases{ 
\begin{cases}
\displaystyle{ {N^2 |k^2|^{\Delta - 2} \Gamma
 (3-\Delta )\epsilon^{2(\Delta -4)}
\over 8\pi^2 2^{2\Delta -5} \Gamma (\Delta -2 )}\,,}
                          &  k^2 > 0,  \cr
           \noalign{\vskip 4pt}
          \displaystyle{ {N^2 |k^2|^{\Delta - 2} \Gamma (3-\Delta )
   \epsilon^{2(\Delta -4)}
\over 8\pi^2 2^{2\Delta -5} \Gamma (\Delta -2 )} \cos{\pi\Delta}\,, }
 &  k^2 < 0,   \cr
\end{cases}
%         }
%   \label{f_grav}
\end{equation}

\begin{equation}
  \Im G^R (k)
 = - {N^2 |k^2|^{\Delta - 2} \Gamma (3-\Delta )\epsilon^{2(\Delta -4)}
\over 8\pi^2 2^{2\Delta -5} \Gamma (\Delta -2 )}
  \sin{(\pi\Delta)} \,  \theta (-k^2)\, \mbox{sgn} \, \omega \,.
\label{green_im}
\end{equation}
\end{subequations}
For the integer values of $\Delta$, we have, correspondingly,
\begin{subequations}
\begin{equation}
 \Re G^R (k) =  { N^2  \epsilon^{2(\Delta - 4 )}
\over  8\pi^2(\Delta - 3)!^2  2^{2\Delta - 5} }
  |k^2|^{\Delta - 2}
\ln{|k^2|}\,, \;\;\;\; \Delta = 2, 3, 4, \ldots
\end{equation}
\begin{equation}
 \Im G^R (k) = - { N^2  \epsilon^{2(\Delta - 4 )}
\over  8\pi (\Delta - 3)!^2  2^{2\Delta - 5} }
  |k^2|^{\Delta - 2}\, \theta (-k^2)\,
  \mbox{sgn}\, \omega  \,, \;\;\;\; \Delta = 2, 3, 4, \ldots
\end{equation}
\end{subequations}
In the massless case ($\Delta = 4$), we get
\begin{equation}
 G^R (k) =  {N^2 k^4\over 64 \pi^2} (\ln{|k^2|} - 
i\pi\,  \theta (-k^2)\, \mbox{sgn} \, \omega)\,.
\end{equation}

\section{Retarded Green's function at high momenta}
\label{sec:high-freq}
In this appendix we consider the retarded Green's function of
$\O=\frac14 F^2$ (the same formulas apply for $\O=\frac14F\tilde F$)
in the regime $\omega,|\k|\gg T$.  Since the temperature now is much
smaller than the momentum scales, one should expect the result to be
close to the zero-temperature result~(\ref{ret_massless}), with the
temperature providing only a small correction.  This can be shown
explicitly by using the Langer-Olver
method~\cite{Policastro:2001yb,Olver_book}.
However, one
qualitative effect occurs at finite temperature: $G^R(k)$ acquires
a small imaginary part for spacelike $k$'s (recall that at zero
temperature, $G^R(k)$ is real if $k^2>0$).  This happens 
for the same reason as the one causing
the Landau damping in plasma: excitations in
the plasma can absorb a spacelike momentum.
We shall estimate $\Im G^R(k)$ for large
spacelike $k$'s.

By a change of variables from $f_k(u)$ to 
$W(u) = u^{-1/2} \sqrt{u^2-1}\, f_k(u)$, the mode equation~(\ref{heun_eq}) 
can be written in the form of the Schr\"{o}dinger equation,
\begin{equation}\label{schroed}
W''(u) = \left( \wn^2 F(u) + G(u)\right) W(u)\,,
\end{equation}
where
\begin{equation}
F(u) = - {1-s^2(1-u^2)\over  u (1-u^2)^2}\,,
\hspace{1cm} G(u) = - {u^4+6 u^2-3\over 4 u^2(1-u^2)^2}\,,
\end{equation}
and $s=|\k|/\omega$.
In the regime, 
$\wn = (\omega R)^2/2\rho_0 = \omega/2\pi T \gg 1$, $s\sim 1$, the term
proportional to $\wn^2$ dominates in the potential, and 
the solution can be found by the WKB approximation.

For $s>1$ (corresponding to timelike $k$'s), the potential in
the Schr\"odinger equation~(\ref{schroed}) is positive for
$u<u_0=\sqrt{1-1/s^2}$ and negative for $u>u_0$.  The solution to
eq.~(\ref{schroed}), thus, decays exponentially in the interval $(0,u_0)$
and oscillates in the interval $(u_0,1)$.  Physically, the particle has
to tunnel from $u=0$ to $u=u_0$ before it can reach the horizon $u=1$.
The imaginary part of $G^R$ is proportional to the tunneling probability,
which is
\begin{equation}
\Im  G^R (\omega, \k) \sim \exp\biggl(-2\wn\!\int\limits_0^{u_0}\!du\,
  \sqrt{F(u)}\biggr)\,.
\end{equation}
For example, if $\omega\ll |\k|$, then
\begin{equation}
  \Im  G^R (\omega, \k) \sim e^{- a |\k|/T}\,, \qquad
  \omega \ll |\k|\,,
\end{equation}
where 
\begin{equation}
  a= \frac{2\Gamma\left(\frac54\right)} 
  {\sqrt\pi\Gamma\left(\frac34\right)} \approx 0.835\,.
\end{equation}
It is not immediately clear how to interpret this value of $a$.  In a
weakly coupled relativistic plasma, an external perturbation with
spatial momentum $(0,|\k|)$ can be absorbed by an particle with
momentum $(|\k|/2,-|\k|/2)$ so that its momentum becomes
$(|\k|/2,|\k|/2)$ and the particle remains on shell.  The probability
of such a process is suppressed by the Boltzmann factor of the
particle, i.e., $e^{-|\k|/(2T)}$.  Since $a>\frac12$, there is some
additional suppression in the strongly coupled CFT.

The formulas presented above are confirmed by the more
sophisticated Langer-Olver method, which is the
WKB approximation for potentials with regular singular points (for more 
information and references see
refs.~\cite{Policastro:2001yb,Olver_book}).

\end{document}